\documentclass[aps,prl,twocolumn,amsmath,amssymb,groupedaddress
]{revtex4-1}
\usepackage{txfonts}
\usepackage{graphicx}
\usepackage{dcolumn}
\usepackage{bm}
\begin{document}
\title{
Magnetic Hedgehog Lattice in a Centrosymmetric Cubic Metal
}
\author{
Shun~Okumura$^1$, Satoru~Hayami$^2$, Yasuyuki~Kato$^3$, and Yukitoshi~Motome$^3$
}
\affiliation{
$^1$The Institute for Solid State Physics, the University of Tokyo, Kashiwa 277-8581, Japan\\
$^2$Faculty of Science, Hokkaido University, Sapporo 060-0810, Japan\\
$^3$Department of Applied Physics, the University of Tokyo, Tokyo 113-8656, Japan
}

\begin{abstract}
The hedgehog lattice (HL) is a three-dimensional topological spin texture hosting a periodic array of magnetic monopoles and antimonopoles.
It has been studied theoretically for noncentrosymmetric systems with the Dzyaloshinskii-Moriya interaction, but the stability, as well as the magnetic and topological properties, remains elusive in the centrosymmetric case.
We here investigate the ground state of an effective spin model with long-range bilinear and biquadratic interactions for a centrosymmetric cubic metal by simulated annealing.
We show that our model stabilizes a HL composed of two pairs of left- and right-handed helices, resulting in no net scalar spin chirality, in stark contrast to the noncentrosymmetric case.
We find that the HL turns into topologically-trivial conical states in an applied magnetic field.
From the detailed analyses of the constituent spin helices, we clarify that the ellipticity and angles of the helical planes change gradually while increasing the magnetic field.
We discuss the results in comparison with the experiments for a centrosymmetric cubic metal SrFeO$_3$.
\end{abstract}

\maketitle

Multiple-$Q$ spin textures, which are superpositions of multiple spin density waves or helices, have attracted much attention in condensed matter physics for several decades~\cite{Bak1978, Bak1980, Forgan1989, Forgan1990, Rossler2006, Martin2008, Muhlbauer2009, Yu2010, Takagi2018, Khanh2022}. 
Of particular interest are the ones hosting periodic arrays of topological objects, typically exemplified by a two-dimensional array of magnetic skyrmions called the skyrmion lattice~\cite{Rossler2006, Muhlbauer2009, Yu2010}. 
The hedgehog lattice (HL) is one of the three-dimensional multiple-$Q$ spin textures given by an array of magnetic hedgehogs and antihedgehogs~\cite{Kanazawa2012, Kanazawa2016}.
The hedgehogs and antihedgehogs can be regarded as magnetic monopoles and antimonopoles, respectively, with respect to the emergent magnetic field arising from the noncoplanar spin configurations through the Berry phase mechanism.
The peculiar distribution of the emergent magnetic field in the HL leads to intriguing macroscopic responses, such as the topological Hall and Nernst effects~\cite{Nagaosa2010, Xiao2010, Kanazawa2011, Hayashi2021, Fujishiro2018}. 

It has been recognized that an antisymmetric exchange interaction, called the Dzyaloshinskii-Moriya (DM) interaction~\cite{Dzyaloshinsky1958, Moriya1960}, plays a crucial role in realizing such multiple-$Q$ spin textures in magnets with noncentrosymmetric crystalline structures~\cite{Rossler2006, Yi2009}.
In recent years, however, a new generation of topological spin textures has been discovered even in the centrosymmetric systems. 
Following theoretical findings of skyrmion lattices stabilized by magnetic frustration in insulators~\cite{Okubo2012, Leonov2015} and spin-charge coupling in metals~\cite{Ozawa2017PRL, Hayami2021}, several candidate substances have been discovered~\cite{Kurumaji2019, Hirshberger2019, Khanh2020, Gao2020}.
HLs have also been observed not only in the noncentrosymmetric $B$20-type compounds MnSi$_x$Ge$_{1-x}$~\cite{Fujishiro2019} but also in the simple cubic perovskite SrFeO$_3$~\cite{Ishiwata2020}.
While the noncentrosymmetric HLs were studied theoretically for the stabilization mechanism and the magnetic and topological properties~\cite{Binz2006PRL, Binz2006PRB, Binz2008, Park2011, Yang2016, Zhang2016, Grytsiuk2020, Okumura2020PRB, Shimizu2021PRB1, Kato2021, Shimizu2021PRB2, Kato2022}, the centrosymmetric ones have not been detailed thus far.

In this Letter, we theoretically study HLs in a centrosymmetric cubic system. 
For an effective spin model with long-range interactions that incorporate the itinerant nature of electrons, we clarify the ground-state phase diagram by using simulated annealing.
The model stabilizes a HL composed of four spin helices ($4Q$-HL) by synergy of bilinear and biquadratic spin interactions.
We show that an external magnetic field causes phase transitions from the topological 4$Q$-HL to three different types of topologically-trivial 4$Q$ conical states ($4Q$-C) depending on the model parameters. 
Analyzing the detailed spin structures of the $4Q$-HL and the $4Q$-C, we clarify how the constituent spin helices evolve with the magnetic field. 
We discuss the results in comparison with the noncentrosymmetric case and the experiments for the centrosymmetric cubic metal SrFeO$_3$~\cite{Ishiwata2011, Ishiwata2020}.

Following the previous study~\cite{Okumura2020PRB}, we consider an effective spin model with long-range interactions arising from the itinerant nature of electrons on a simple cubic lattice; 
we omit the DM-type interaction since we consider a centrosymmetric case in this study. 
The Hamiltonian is given by~\cite{footnote1} 
\begin{eqnarray}
	\mathcal{H} = 2\sum_{\eta}\left[-J\mathbf{S}_{\mathbf{Q}_\eta}\cdot\mathbf{S}_{-\mathbf{Q}_\eta}+\dfrac{K}{N}({\mathbf S}_{\mathbf{Q}_\eta}\cdot{\mathbf S}_{-\mathbf{Q}_\eta})^2\right]-\sum_{l}\mathbf{h}\cdot\mathbf{S}_{\mathbf{r}_l},
\label{eq: Heff}
\end{eqnarray}
where $\mathbf{S}_{\mathbf{q}} = \frac{1}{\sqrt{N}} \sum_l \mathbf{S}_{\mathbf{r}_l} e^{i\mathbf{q}\cdot\mathbf{r}_l}$, $\mathbf{S}_{\mathbf{r}_l}$ represents the spin at site $l$, $\mathbf{r}_l$ is the position vector of the site $l$, and $N$ is the number of spins. 
The first term denotes the bilinear interaction called the Ruderman-Kittel-Kasuya-Yosida interaction~\cite{Ruderman1954, Kasuya1956, Yosida1957}; we set the energy scale as $J=1$. 
The second term is the biquadratic interaction with a positive coupling constant $K>0$, which is the most relevant term among the higher-order perturbations with respect to the spin-charge coupling~\cite{Akagi2012, Hayami2014, Hayami2017}. 
The sum is taken for a set of the tetrahedral wave vectors as $\mathbf{Q}_1=(Q, -Q, -Q)$, $\mathbf{Q}_2=(-Q, Q, -Q)$, $\mathbf{Q}_3=(-Q, -Q, Q)$, and $\mathbf{Q}_4=(Q, Q, Q)$, following the previous model for the $4Q$-HL~\cite{Okumura2020PRB}.
The last term in Eq.~(\ref{eq: Heff}) describes the Zeeman coupling to an external magnetic field $\mathbf{h}$. 
Note that the energy of the model in Eq.~\eqref{eq: Heff} is independent of the field direction because of the absence of the spin anisotropy.
In the following calculations, we take $\bold{h}=\frac{1}{\sqrt{3}}(h,h,h)\parallel\bold{Q}_4$, treat the spins as classical vectors with $|\mathbf{S}_{\mathbf{r}_l}|=1$ for simplicity, and take $Q=\pi/8$ in the system with $N=16^3$ under periodic boundary conditions, for which the finite-size effect is negligible.

We study the ground state of the model in Eq.~\eqref{eq: Heff} by simulated annealing with gradually reducing temperature from $T=1$ to $T=10^{-5}$ with a condition $T_n=10^{-0.1n}$, where $T_n$ is the temperature in the $n$th step.
We spend a total of $10^5-10^6$ Monte Carlo sweeps during the annealing by using the standard Metropolis algorithm.
After annealing at a set of $K$ and $h$, we increase or decrease $K$ and $h$ successively by $\Delta K=0.02$ and $\Delta h= 0.05$, respectively.
At every shift by $\Delta K$ or $\Delta h$, we heat the system up to $T=10^{-3}$ and cool it down again to $T=10^{-5}$ by the same scheme of annealing. 
Carefully comparing the results by starting from various $K$ and $h$, we obtain the lowest-energy state at each $K$ and $h$. 
To identify the magnetic phases, we calculate the magnetic moment with wave vector $\mathbf{q}$, $m_\mathbf{q}=\sqrt{S(\mathbf{q})/N}$, where $S(\mathbf{q})$ is the spin structure factor defined by $S(\mathbf{q})=\frac{1}{N}\sum_{l,l'}\mathbf{S}_{\mathbf{r}_l}\cdot\mathbf{S}_{\mathbf{r}_{l'}}e^{i\mathbf{q}\cdot(\mathbf{r}_{l}-\mathbf{r}_{l'})}$; $m_{\mathbf{q}=0}$ corresponds to the magnetization per spin.
We also compute the number of hedgehog-antihedgehog pairs, $N_\mathrm{pair} = \sum_j |Q_\mathrm{m}(\bold{r}_j)|/2$, where the sum is taken for the magnetic unit cell, and $Q_\mathrm{m}(\bold{r}_j)$ is the topological number called the monopole charge~\cite{Park2011, Okumura2020PRB, Okumura2020JPSCP}; $Q_\mathrm{m}(\bold{r}_j)$ takes the value of $+1$ $(-1)$ when a (anti)monopole exists in a $j$th unit cube at $\bold{r}_j$.

\begin{figure}[t]
\centering
\includegraphics[width=\columnwidth,clip]{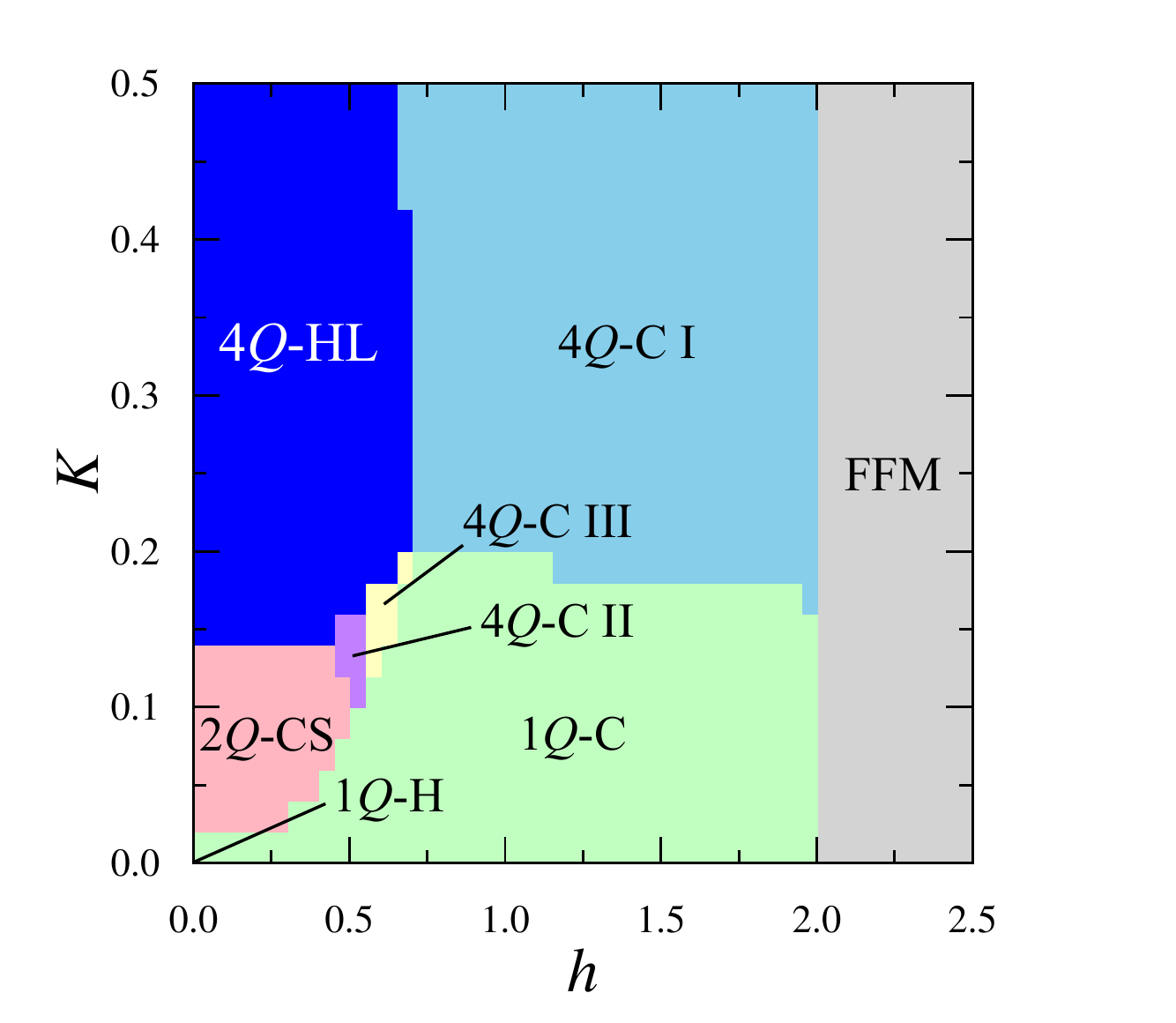}
\caption{
Phase diagram of the model in Eq.~(\ref{eq: Heff}) for the magnetic field $h$ and the biquadratic interaction $K$.
4$Q$-HL, $4Q$-C I--III, 2$Q$-CS, 1$Q$-H, 1$Q$-C, and FFM represent the 4$Q$ hedgehog lattice, the three different 4$Q$ conical states, the 2$Q$ chiral stripe, the 1$Q$ helical state, the 1$Q$ conical state, and the forced ferromagnetic state, respectively.
}
\label{f1}
\end{figure}

\begin{figure*}[t]
\centering
\includegraphics[width=\linewidth,clip]{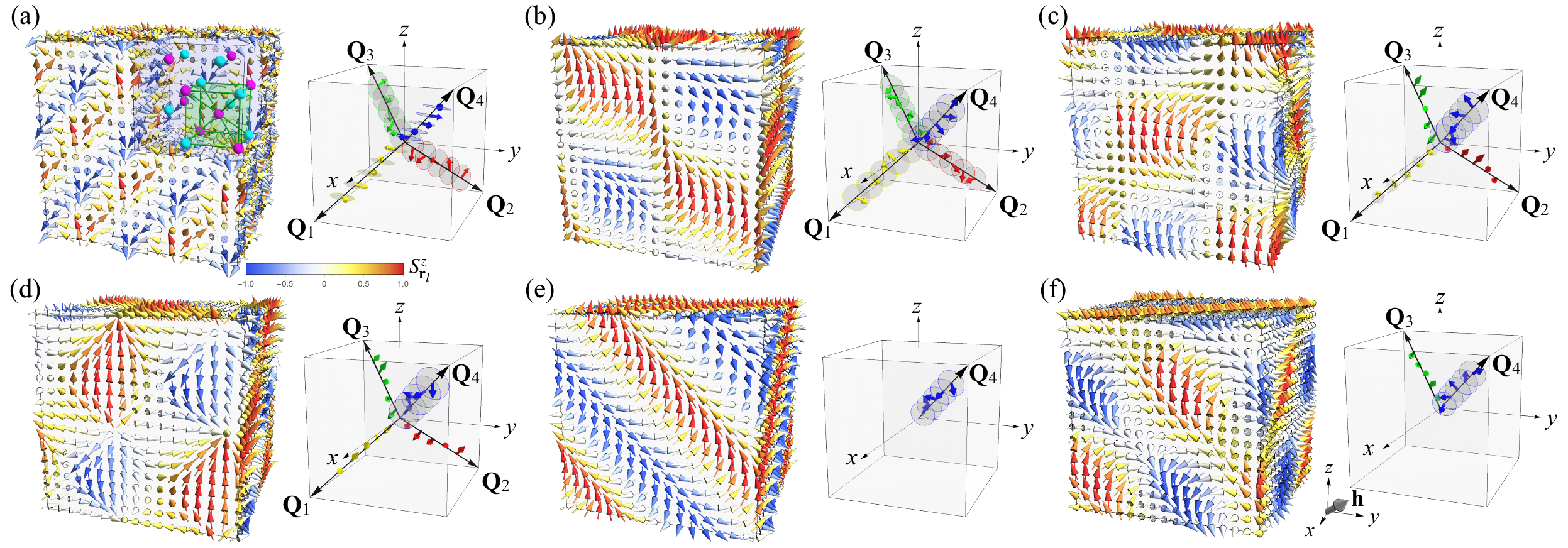}
\caption{
Spin textures and constituent waves obtained by simulated annealing for the model in Eq.~(\ref{eq: Heff}): (a) 4$Q$-HL at $K=0.3$ and $h=0$, (b) 4$Q$-C I at $K=0.3$ and $h=1$, (c) 4$Q$-C II at $K=0.14$ and $h=0.5$, (d) 4$Q$-C III at $K=0.14$ and $h=0.6$, (e) 1$Q$-C at $K=0.06$ and $h=0.7$, and (f) 2$Q$-CS at $K=0.06$ and $h=0.2$.
The color of the arrows in the left panels denotes the $z$ component of the spins, $S^z_{\bold{r}_l}$, as indicated in the inset of (a).
In (a), there are eight hedgehogs (magenta) and eight antihedgehogs (cyan) in the bluish magnetic unit cell, which form two interpenetrating body-centered-cubic lattices as shown by the green guides.
The gray arrow in the inset of (f) represents the magnetic field along the [111] direction, $\bold{h}$.
The yellow, red, green, and blue arrows in the right panels represent the constituent waves with $\bold{Q}_1$, $\bold{Q}_2$, $\bold{Q}_3$, and $\bold{Q}_4$, and the gray circles show the helical plane of each helix.
}
\label{f2}
\end{figure*}

Figure~\ref{f1} shows the phase diagram obtained by the simulated annealing.
At zero field, the bilinear interaction stabilizes the single-$Q$ helical state ($1Q$-H) of any $\bold{Q}_\eta$ at $K=0$. 
When introducing $K$, the double-$Q$ chiral stripe ($2Q$-CS) appears for $0<K\lesssim0.13$, which is a superposition of a helix and a sinusoid propagating in different directions (any two of $\bold{Q}_\eta$)~\cite{Ozawa2016}.
For $K\gtrsim0.13$, the system stabilizes the $4Q$-HL, whose spin texture is displayed in Fig.~\ref{f2}(a).
The spin configuration is composed of a superposition of two pairs of spin helices whose amplitudes are the same but helical axes are orthogonal to each other, as approximately given by
\begin{align}
	\bold{S}_{\bold{r}_l}\propto\left(\sum_{\eta=1,4}c_{\eta,l}+\sum_{\eta=2,3}\chi_\eta\xi_\eta s_{\eta,l}, \sum_{\eta=1,4}\chi_\eta\xi_\eta s_{\eta,l}, \sum_{\eta=2,3}c_{\eta,l}\right),
\label{eq: 4QHL}
\end{align}
where $c_{\eta,l} = \cos(\mathcal{Q}_{\eta,l})$, $s_{\eta,l}=\sin(\mathcal{Q}_{\eta,l})$, $\mathcal{Q}_{\eta,l}=\bold{Q}_\eta\cdot\bold{r}_l+\varphi_\eta$ with the phase degrees of freedom $\varphi_\eta$, $\xi_\eta$ represents the ellipticity of each helical plane ($0<\xi_\eta<1$), and the chirality of each helix, $\chi_\eta$, takes a value of $+1$ or $-1$ for a right- or left-handed helix; 
any combinations of $\eta$ are allowed, and any global spin rotation is also allowed.
We find that $\xi_\eta$ is common to the four components and the value decreases with increasing $K$. 
We also find that the values of $\chi_\eta$ appear in pairs; we will discuss its implication later.
In the example in Fig.~\ref{f2}(a), $\chi_\eta$ takes $-1$ for $\eta=1,2$ and $+1$ for $\eta=3,4$.
In addition, we note that $\sum_{\eta}\varphi_\eta=\pi/3$~\cite{Shimizu2022PRB}.
Similar to the noncentrosymmetric case~\cite{Okumura2020PRB}, this $4Q$-HL state accommodates eight pairs of hedgehogs and antihedgehogs, i.e., $N_\mathrm{pair}=8$, forming two interpenetrating body-centered-cubic structures, as shown in Fig.~\ref{f2}(a).
We note that the phase boundary between the $2Q$-CS and $4Q$-HL locates at lower $K$ than the previous result in the absence of the DM-type interaction obtained by variational calculations~\cite{footnote1, Okumura2020PRB}.

By applying the magnetic field $h$, the $4Q$-HL exhibits phase transitions to several topologically trivial states without hedgehogs and antihedgehogs, as shown in Fig.~\ref{f1}: three types of the quadruple-$Q$ conical states ($4Q$-C I, II, and III), the single-$Q$ conical state ($1Q$-C) generated by spin canting from the $1Q$-H at $K=h=0$, and the forced ferromagnetic state (FFM) for $h>2$.  
The typical spin configuration in each phase including the $2Q$-CS is displayed in Figs.~\ref{f2}(b)--\ref{f2}(f) (except for the FFM).  
All the above states are described by a superposition of helices and sinusoids in a unified form as
\begin{align}
	\bold{S}_{\bold{r}_l}\propto\sum_{\eta}a_\eta\left[c_{\eta,l}\bold{e}^1+\chi_\eta s_{\eta,l}\bold{e}^2\right]+\left[\sum_{\eta}b_\eta s_{\eta,l}+m\right]\bold{e}^0,
\label{eq: conical}
\end{align}
where $\bold{e}^0$ is the unit vector parallel to the field direction, $\bold{e}^1$ and $\bold{e}^2$ are the other unit vectors satisfying $\bold{e}^0=\bold{e}^1\times\bold{e}^2$, and $m$ is the uniform magnetization; $a_\eta$ and $b_\eta$ are amplitudes of the helices and sinusoids, respectively.
The $4Q$-C I in Fig.~\ref{f2}(b) is given by a superposition of four helices with the same amplitudes, namely, $^\forall a_\eta=a\neq 0$ and $^\forall b_\eta=0$, whose chirality are paired in two similar to the $4Q$-HL; $\chi_{1,2}=-1$ and $\chi_{3,4}=+1$ in Fig.~\ref{f2}(b). 
Meanwhile, the $4Q$-C II in Fig.~\ref{f2}(c) consists of two helices with different amplitudes and two sinusoids with the same amplitudes, namely, $0<a_1<b_2=b_3<a_4$ and the other $a_\eta$ and $b_\eta$ are zero, and the two helices have opposite chirality [$\chi_1=+1$ and $\chi_4=-1$ in Fig.~\ref{f2}(c)].
The $4Q$-C III in Fig.~\ref{f2}(d) is composed of a helix and three sinusoids with $0<b_1=b_2=b_3<a_4$ and $\chi_4$ is either $+1$ or $-1$.
Finally, the $1Q$-C in Fig.~\ref{f2}(e) and  the $2Q$-CS in Fig.~\ref{f2}(f) are characterized by $a_4=1$ and $0<b_3<a_4$, respectively; $\chi_4$ is also either $+1$ or $-1$ in both cases.
In the $4Q$-C I--III, $1Q$-C, and $2Q$-CS, all the helices are perfectly circular ($\xi_\eta=1$) with the helical axes parallel to the field direction, and all the sinusoids oscillate in the field direction.
Note that for all the above states any combinations of $\eta$ are energetically degenerate in the current isotropic model. 

\begin{figure}[t]
\centering
\includegraphics[width=\columnwidth,clip]{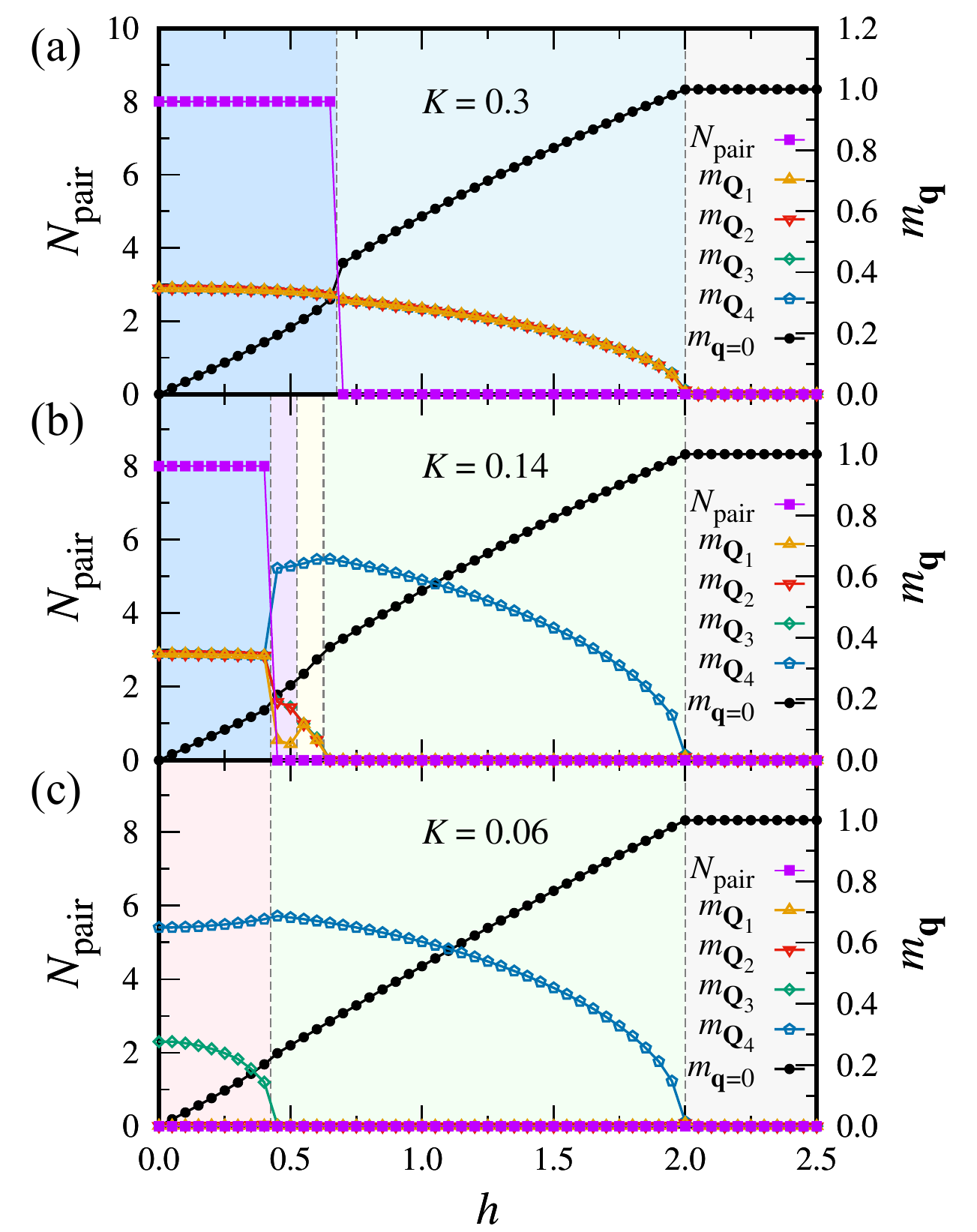}
\caption{
Magnetic field dependences of the number of hedgehog-antihedgehog pairs, $N_{\rm pair}$, and the magnetic moments with wave vectors, $m_{\mathbf{q}}$, at (a) $K=0.3$, (b) $K=0.14$, and (c) $K=0.06$.
$m_{\mathbf{q}=0}$ represents the magnetization.
The vertical dashed lines are the critical magnetic fields, and the background colors represent the phases in Fig.~\ref{f2}.
}
\label{f3}
\end{figure}

Figure~\ref{f3} shows the magnetic field dependences of $N_{\rm pair}$ and $m_{\bold{q}}$ at three different values of $K$.
At $K = 0.3$ in Fig.~\ref{f3}(a), the $4Q$-HL with $N_\mathrm{pair}=8$ turns into the $4Q$-C I with $N_\mathrm{pair}=0$ at $h\simeq0.675$.
The magnetization $m_{\bold{q}=0}$ increases with $h$ and jumps at $h\simeq0.675$, while $m_{\bold{Q}_\eta}$ decrease equally and also jump at $h\simeq0.675$. 
These indicate that the phase transition is of the first order. 
Meanwhile, at $K=0.14$ in Fig.~\ref{f3}(b), the $4Q$-HL turns into the $4Q$-C II with a similar change of $N_{\rm pair}$ at $h\simeq0.425$, and the $4Q$-C II changes into the $4Q$-C III at $h\simeq0.525$. 
Both are first-order transitions resulting from the changes of the types of constituent waves; see Figs.~\ref{f2}(a), \ref{f2}(c) and \ref{f2}(d).
While further increasing $h$, the $4Q$-C III turns into the $1Q$-C at $h\simeq0.625$, with continuous changes $m_{\bold{Q}_{\eta\neq4}}\rightarrow0$, namely, $b_{\eta\neq4}\rightarrow0$ in Eq.~\eqref{eq: conical}.
At $K=0.06$ in Fig.~\ref{f3}(c), the $2Q$-CS turns into the $1Q$-C at $h\simeq0.425$ with vanishing sinusoidal component $m_{\bold{Q}_{\eta\neq4}}\to 0$, namely, $b_{\eta\neq4}\to 0$ in Eq.~\eqref{eq: conical}.
In all the cases, the transitions to the FFM at $h = 2$ are continuous.

\begin{figure}[t]
\centering
\includegraphics[width=\columnwidth,clip]{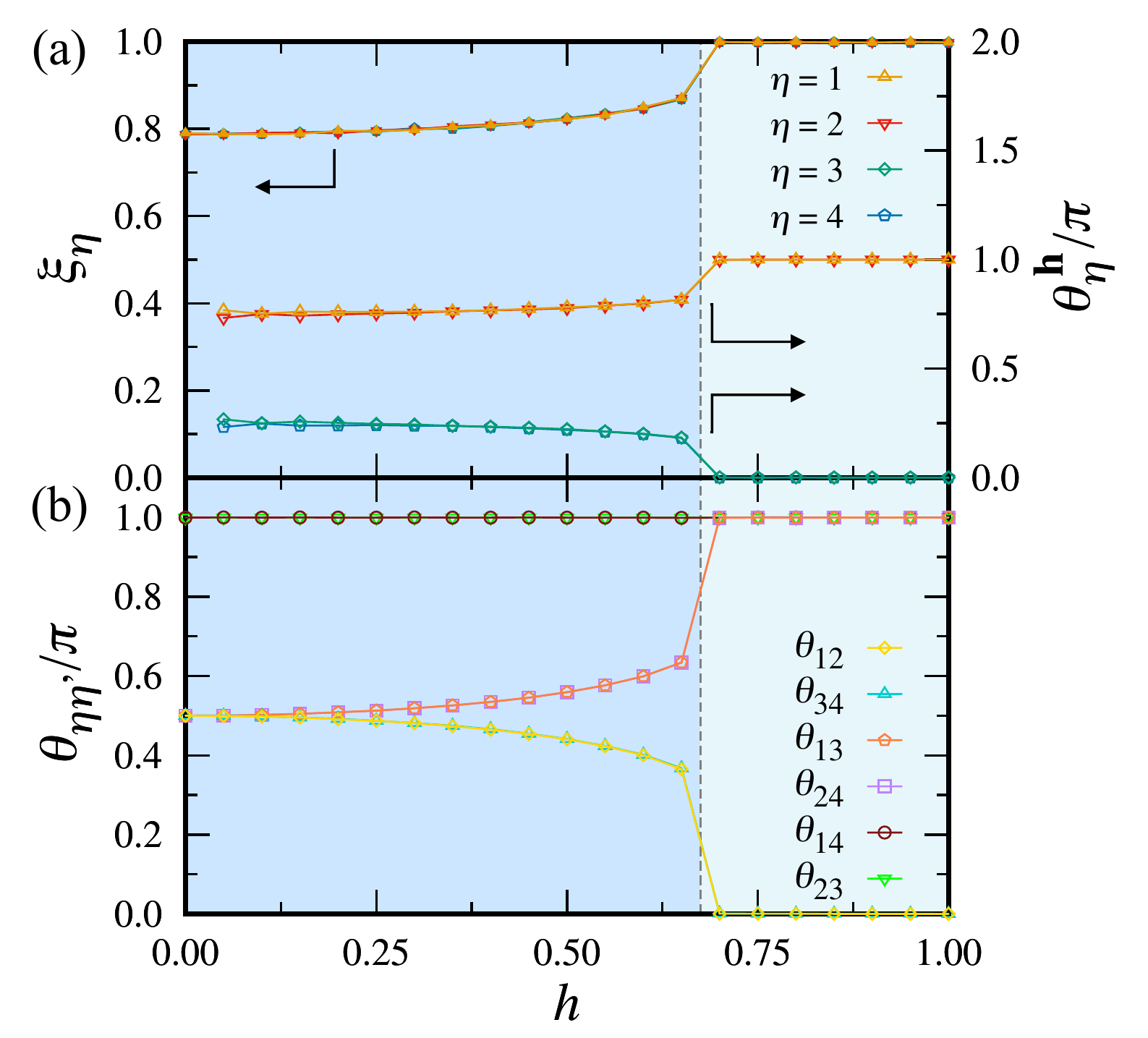}
\caption{
Magnetic field dependences of (a) the ellipticity of the helical plane, $\xi_\eta$, the angles between $\bold{h}$ and the helical axis $\bold{u}^0_\eta$, $\theta^{\bold{h}}_{\eta}$, and (b) the relative angles between the helical axes, $\theta_{\eta\eta'}$, at $K=0.3$.
}
\label{f4}
\end{figure}

Let us closely look into the field dependence of the magnetic structure of the $4Q$-HL, focusing on the case with $K=0.3$.
Using the spin configurations obtained by simulated annealing, we obtain the ellipticity of each helical plane $\xi_\eta$ and the direction of the helical axis $\bold{u}^0_\eta$ ($|\bold{u}^0_\eta|=1$)~\cite{footnote2}.
Figure~\ref{f4}(a) shows the changes of $\xi_\eta$ and  the angles between the helical axes and the field direction, $\theta^{\bold{h}}_\eta=\arccos\left(\bold{u}^0_\eta\cdot\bold{e}^0\right)$. 
We find that, while increasing $h$, all $\xi_\eta$ increase equally and jump to $1$ (perfectly circular) at the first-order transition to the $4Q$-C I at $h\simeq0.675$. 
Meanwhile, $\theta_\eta$ are grouped into two: 
$\theta^{\bold{h}}_1$ and $\theta^{\bold{h}}_2$ increase gradually from $\sim \frac{3\pi}{4}$ with $h$ and jump to $\pi$ at $h\simeq0.675$, while $\theta^{\bold{h}}_3$ and $\theta^{\bold{h}}_4$ decrease from $\sim \frac{\pi}{4}$ and finally vanish.
This indicates that the helical axes for $\eta=3, 4$ ($1, 2$) are gradually tilted to (away from) the magnetic field direction, and become (anti)parallel to the magnetic field at the first-order transition.

Figure~\ref{f4}(b) shows the relative angles between the helical axes, $\theta_{\eta\eta'}=\arccos\left(\bold{u}^0_\eta\cdot\bold{u}^0_{\eta'}\right)$.
We find that $\theta_{12} = \theta_{34} = \theta_{13} = \theta_{24} = \frac{\pi}{2}$ at zero field: 
the corresponding $\bold{u}^0_\eta$ are orthogonal to each other. 
While increasing $h$, these $\theta_{\eta\eta'}$ change gradually in pairs, and finally, $\theta_{12}$ and $\theta_{34}$ ($\theta_{13}$ and $\theta_{24}$) become $0$ ($\pi$): 
$\bold{u}^0_1$ and $\bold{u}^0_2$, $\bold{u}^0_3$ and $\bold{u}^0_4$ ($\bold{u}^0_1$ and $\bold{u}^0_3$, $\bold{u}^0_2$ and $\bold{u}^0_4$) become (anti)parallel in the $4Q$-I phase for $h\gtrsim 0.675$. 
Interestingly, $\theta_{14} = \theta_{23} = \pi$ for all $h$; namely the helical axis $\bold{u}^0_{1(2)}$ is always antiparallel to $\bold{u}^0_{4(3)}$.
This indicates that the chirality $\chi_{1(2)}$ has the opposite sign to $\chi_{4(3)}$: the right-handed and left-handed helices appear in pairs. 
Therefore, the net scalar spin chirality is always zero, even in an applied magnetic field, suggesting no topological Hall effect. 
We confirm it by directly computing the scalar spin chirality. 
Furthermore, we find no movement of the hedgehogs and antihedgehogs by the magnetic field, and hence, no topological transition due to their pair annihilation.
These aspects are in stark contrast to the noncentrosymmetric cases with the DM interaction~\cite{Okumura2020PRB}.

Finally, let us discuss our results in comparison with the experimental data for SrFeO$_3$. 
This material has a centrosymmetric cubic lattice and exhibits not only a single-$Q$ helical state~\cite{MacChesney1965, Takeda1972, Oda1977} but also multiple-$Q$ magnetic phases~\cite{Ishiwata2011, Reehuis2012, Ishiwata2020}.
The 4$Q$-HL in our results for large $K$ appears to be related with the quadruple-$Q$ phase in SrFeO$_3$ at finite temperature~\cite{Ishiwata2020}, considering that the biquadratic interaction $K$ can be effectively enhanced by raising temperature~\cite{Reimers1991, Okubo2011}.
The field-induced transition from the $4Q$-HL to the $1Q$-C in Fig.~\ref{f3}(b) is also consistent with the experimental observation~\cite{Ishiwata2020}.
In addition, the $2Q$-CS in the small-$K$ region appears to be relevant to the double-$Q$ phase in SrFeO$_3$ at low temperature~\cite{Ishiwata2020, Yambe2020}, which changes into the single-$Q$ phase~\cite{Mostovoy2005, Azhar2017} with decreasing the magnetic moments not parallel to the magnetic field similarly to the result in Fig.~\ref{f3}(c).
However, our results suggest no topological Hall effect, in contrast to the experiments, in both double- and quadruple-$Q$ phases in SrFeO$_3$~\cite{Hayashi2001, Ishiwata2011}.
We speculate that the discrepancy can be reconciled, for instance, by introducing magnetic anisotropy that differentiates the amplitudes, angles, and ellipticity of helical planes~\cite{Shimizu2021PRB1, Kato2022}; we will discuss the effects of cubic single-ion anisotropy elsewhere.

In summary, we have numerically demonstrated that the synergy between the long-range bilinear and biquadratic interactions leads to a variety of multiple-$Q$ spin textures, including the $4Q$-HL, even in a centrosymmetric system.
We showed that the $4Q$-HL at zero field consists of two pairs of elliptically distorted spin helices whose helical axes are orthogonal to each other, while the angles and ellipticity are changed gradually by the external magnetic field before entering the $4Q$-C. 
We also found that one of the $4Q$-C states consists of four spin helices like the $4Q$-HL, whereas the rest two are composed of mixtures of helices and sinusoids.
These behaviors of the $4Q$-HL and $4Q$-C states in the magnetic field are qualitatively different from those in the noncentrosymmetric systems in the presence of the DM interaction~\cite{Okumura2020PRB}. 
Notably, we found that our centrosymmetric model exhibits no net scalar spin chirality, which is a source of the topological effect.
While we have studied the ground state only, it is important to study the effects of temperature~\cite{Kato2022} and magnetic anisotropy for understanding of the experiments.
It is also an interesting issue to explore characteristic phenomena in the centrosymmetric HL, such as magnetic excitations~\cite{Kato2021}, transport, and optical responses. 

\begin{acknowledgments}
We would like to thank K.~Aoyama, R.~Eto, M.~Gen, M.~Mochizuki, K.~Shimizu, and R.~Yambe for fruitful discussions.
This research was supported by JST CREST (Nos.~JPMJCR18T2 and JPMJCR19T3), JST PRESTO (No.~JPMJPR20L8), JSPS KAKENHI (Nos.~JP19H05825, JP21H01037, JP22H04468, JP22K03509, and JP22K13998), the Chirality Research Center in Hiroshima University, and JSPS Core-to-Core Program, Advanced Research Networks.
This work was also supported by ``Joint Usage/Research Center for Interdisciplinary Large-scale Information Infrastructures" and ``High Performance Computing Infrastructure" in Japan (Project ID: EX20304).
Parts of the numerical calculations were performed in the supercomputing systems in ISSP, the University of Tokyo.
SO was supported by JSPS through the research fellowship for young scientists.
\end{acknowledgments}


\begin{thebibliography}{99}

\bibitem{Bak1978} P.~Bak and B.~Lebech, Phys. Rev. Lett. {\bf 40}, 800 (1978). 
\bibitem{Bak1980} P.~Bak and M.~H.~Jensen, J. Phys. C: Solid State Phys. {\bf13} L881 (1980). 
\bibitem{Forgan1989} E.~M.~Forgan, E.~P.~Gibbons, K.~A.~McEwen, and D.~Fort, Phys. Rev. Lett. {\bf 62}, 470 (1989). 
\bibitem{Forgan1990} E.~M.~Forgan, B.~D.~Rainford, S.~L.~Lee, J.~S. Abell, and Y.~Bi, J. Phys.: Condens. Matter. {\bf 2}, 10211 (1990). \bibitem{Rossler2006} U.~K.~R\"{o}\ss ler, A.~N.~Bogdanov, and C.~Pfleiderer, Nature {\bf 442}, 797 (2006). 
\bibitem{Martin2008} I.~Martin and C.~D.~Batista, Phys. Rev. Lett. {\bf 101}, 156402 (2008). 
\bibitem{Muhlbauer2009} S.~M\"{u}hlbauer, B.~Binz, F.~Jonietz, C.~Pfleiderer, A.~Rosch, A.~Neubauer, R.~Georgii, and P.~B\"{o}ni, Science {\bf 323}, 915 (2009). 
\bibitem{Yu2010} X.~Z.~Yu, Y.~Onose, N.~Kanazawa, J.~H.~Park, J.~H.~Han, Y.~Matsui, N.~Nagaosa, and Y.~Tokura, Nature {\bf 465}, 901 (2010). 
\bibitem{Takagi2018} R.~Takagi, J.~S.~White, S.~Hayami, R.~Arita, D.~Honecker, H.~M.~R\o nnow, Y.~Tokura, S.~and Seki, Sci. Adv. {\bf 4}, eaau3402 (2018). 
\bibitem{Khanh2022} N.~D.~Khanh, T.~Nakajima, S.~Hayami, S.~Gao, Y.~Yamasaki, H.~Sagayama, H.~Nakao, R.~Takagi, Y.~Motome, Y.~Tokura, T.~Arima, and S.~Seki, Adv. Sci. {\bf 9}, 2105452 (2022).
\bibitem{Kanazawa2012} N.~Kanazawa, J.-H.~Kim, D.~S.~Inosov, J.~S.~White, N.~Egetenmeyer, J.~L.~Gavilano, S.~Ishiwata, Y.~Onose, T.~Arima, B.~Keimer, and Y.~Tokura, Phys. Rev. B {\bf 86}, 134425 (2012).
\bibitem{Kanazawa2016} N.~Kanazawa, Y.~Nii, X.~X.~Zhang, A.~S.~Mishchenko, G.~D.~Filippis, F.~Kagawa, Y.~Iwasa, N.~Nagaosa, and Y.~Tokura, Nat. Commun. {\bf 7}, 11622 (2016).
\bibitem{Nagaosa2010} N.~Nagaosa, J.~Sinova, S.~Onoda, A.~H.~MacDonald, and N.~P.~Ong, Rev. Mod. Phys. {\bf 82}, 1539 (2010).
\bibitem{Xiao2010} D.~Xiao, M.-C.~Chang, and Q.~Niu, Rev. Mod. Phys. {\bf 82}, 1959 (2010).
\bibitem{Kanazawa2011} N.~Kanazawa, Y.~Onose, T.~Arima, D.~Okuyama, K.~Ohoyama, S.~Wakimoto, K.~Kakurai, S.~Ishiwata, and
Y.~Tokura, Phys. Rev. Lett. {\bf 106}, 156603 (2011).
\bibitem{Fujishiro2018} Y.~Fujishiro, N.~Kanazawa, T.~Shimojima, A.~Nakamura, K.~Ishizaka, T.~Koretsune, R.~Arita, A.~Miyake, H. ~Mitamura, K.~Akiba, M.~Tokunaga, J.~Shiogai, S.~Kimura, S.~Awaji, A.~Tsukazaki, A.~Kikkawa, Y.~Taguchi, and Y.~Tokura, Nat. Commun. {\bf 9}, 408 (2018).
\bibitem{Hayashi2021} Y.~Hayashi, Y.~Okamura, N.~Kanazawa, T.~Yu, T.~Koretsune, R.~Arita, A.~Tsukazaki, M.~Ichikawa, M.~Kawasaki, Y.~Tokura, and Y.~Takahashi, Nat. Commun. {\bf 12}, 5974 (2021).
\bibitem{Dzyaloshinsky1958} I.~Dzyaloshinsky, J. Phys. Chem. Solids {\bf 4}, 241 (1958).
\bibitem{Moriya1960} T.~Moriya, Phys. Rev. {\bf 120}, 91 (1960).
\bibitem{Yi2009} S.~D.~Yi, S.~Onoda, N.~Nagaosa, and J.~H.~Han, Phys. Rev. B {\bf 80}, 054416 (2009).
\bibitem{Okubo2012} T.~Okubo, S.~Chung, and H.~Kawamura, Phys. Rev. Lett. {\bf 108}, 017206 (2012).
\bibitem{Leonov2015} A.~O.~Leonov and M.~Mostovoy, Nat. Commun. {\bf 6}, 8275 (2015).
\bibitem{Ozawa2017PRL} R.~Ozawa, S.~Hayami, and Y.~Motome, Phys. Rev. Lett. {\bf 118}, 147205 (2017).
\bibitem{Hayami2021}S.~Hayami and Y.~Motome, J. Phys.: Condens. Matter {\bf 33}, 443001 (2021).
\bibitem{Kurumaji2019} T.~Kurumaji, T.~Nakajima, M.~Hirschberger, A.~Kikkawa, Y.~Yamasaki, H.~Sagayama, H.~Nakao, Y.~Taguchi, T.~Arima, and Y.~Tokura, Science {\bf 365}, 914 (2019).
\bibitem{Hirshberger2019} M.~Hirschberger, T.~Nakajima, S.~Gao, L.~Peng, A.~Kikkawa, T.~Kurumaji, M.~Kriener, Y.~Yamasaki, H.~Sagayama, H.~Nakao, K.~Ohishi, K.~Kakurai, Y.~Taguchi, X.~Yu, T.~Arima, and Y.~Tokura, Nat. Commun. {\bf 10}, 5831 (2019).
\bibitem{Khanh2020} N.~D.~Khanh, T.~Nakajima, X.~Yu, S.~Gao, K.~Shibata, M.~Hirschberger, Y.~Yamasaki, H.~Sagayama, H.~Nakao, L.~Peng, K.~Nakajima, R.~Takagi, T.~Arima, Y.~Tokura, and S.~Seki, Nat. Nanotech. {\bf 15}, 444 (2020).
\bibitem{Gao2020} S.~Gao, H.~D.~Rosales, F.~A.~G.~Albarracín, V.~Tsurkan, G.~Kaur, T.~Fennell, P.~Steffens, M.~Boehm, P.~\v{C}erm\'{a}k, A.~Schneidewind, E.~Ressouche, D.~C.~Cabra, C.~R\"{u}egg, and O.~Zaharko, Nature {\bf 586} 37 (2020).
\bibitem{Fujishiro2019} Y.~Fujishiro, N.~Kanazawa, T.~Nakajima, X.~Z.~Yu, K.~Ohishi, Y.~Kawamura, K.~Kakurai, T.~Arima, H.~Mitamura, A.~Miyake, K.~Akiba, M.~Tokunaga, A.~Matsuo, K.~Kindo, T.~Koretsune, R.~Arita, and Y.~Tokura, Nat. Commun. {\bf 10}, 1059 (2019).
\bibitem{Ishiwata2020} S.~Ishiwata, T.~Nakajima, J.-H.~Kim, D.~S.~Inosov, N.~Kanazawa, J.~S.~White, J.~L.~Gavilano, R.~Georgii, K.~M.~Seemann, G.~Brandl, P.~Manuel, D.~D.~Khalyavin, S.~Seki, Y.~Tokunaga, M.~Kinoshita, Y.~W.~Long, Y.~Kaneko, Y.~Taguchi, T.~Arima, B.~Keimer, and Y.~Tokura, Phys. Rev. B {\bf 101}, 134406 (2020).
\bibitem{Binz2006PRL} B.~Binz, A.~Vishwanath, and V.~Aji, Phys. Rev. Lett. {\bf 96}, 207202 (2006).
\bibitem{Binz2006PRB} B.~Binz and A.~Vishwanath, Phys. Rev. B {\bf 74}, 214408 (2006).
\bibitem{Park2011} J.-H.~Park and J.~H.~Han, Phys. Rev. B {\bf 83}, 184406 (2011).
\bibitem{Yang2016} S.-G.~Yang, Y.-H.~Liu, and J.~H.~Han, Phys. Rev. B {\bf 94}, 054420 (2016).
\bibitem{Grytsiuk2020} S.~Grytsiuk, J.-P.~Hanke, M.~Hoffmann, J.~Bouaziz, O.~Gomonay, G.~Bihlmayer, Y.~Mokrousov, and S.~Blügel, Nat. Commun. {\bf 11}, 511 (2020).
\bibitem{Okumura2020PRB} S.~Okumura, S.~Hayami, Y.~Kato, and Y.~Motome, Phys. Rev. B {\bf 101}, 144416 (2020).
\bibitem{Shimizu2021PRB1} K.~Shimizu, S.~Okumura, Y.~Kato, and Y.~Motome, Phys. Rev. B {\bf 103}, 054427 (2021).
\bibitem{Kato2022} Y.~Kato and Y.Motome, Phys. Rev. B {\bf 105}, 174413 (2022).
\bibitem{Binz2008} B.~Binz and A.~Vishwanath, Physica B {\bf 403}, 1336 (2008).
\bibitem{Zhang2016} X.-X.~Zhang, A.~S.~Mishchenko, G.~De~Filippis, and N.~Nagaosa, Phys. Rev. B {\bf 94}, 174428 (2016).
\bibitem{Kato2021} Y.~Kato, S.~Hayami, and Y.~Motome, Phys. Rev. B {\bf 104}, 224405 (2021).
\bibitem{Shimizu2021PRB2} K.~Shimizu, S.~Okumura, Y.~Kato, and Y.~Motome, Phys. Rev. B {\bf 103}, 184421 (2021).
\bibitem{Ishiwata2011} S.~Ishiwata, M.~Tokunaga, Y.~Kaneko, D.~Okuyama, Y.~Tokunaga, S.~Wakimoto, K.~Kakurai, T.~Arima, Y.~Taguchi, and Y.~Tokura, Phys. Rev. B. {\bf 84}, 054427 (2011).
\bibitem{footnote1}  In the previous study~\cite{Okumura2020PRB}, the factor of 2 in Eq.~\eqref{eq: Heff} was missing from the Hamiltonian, but the calculations included it.
\bibitem{Ruderman1954} M.~A.~Ruderman and C.~Kittel, Phys. Rev. {\bf 96}, 99 (1954).
\bibitem{Kasuya1956} T.~Kasuya, Prog. Theor. Phys. {\bf 16}, 45 (1956).
\bibitem{Yosida1957} K.~Yosida, Phys. Rev. {\bf 106}, 893 (1957).
\bibitem{Akagi2012} Y.~ Akagi, M.~Udagawa, and Y.~Motome, Phys. Rev. Lett. {\bf 108}, 096401 (2012).
\bibitem{Hayami2014} S.~Hayami and Y.~Motome, Phys. Rev. B {\bf 90}, 060402(R) (2014).
\bibitem{Hayami2017} S.~Hayami, R.~Ozawa, and Y.~Motome, Phys. Rev. B {\bf 95}, 224424 (2017).
\bibitem{Okumura2020JPSCP} S.~Okumura, S.~Hayami, Y.~Kato, and Y.~Motome, JPS Conf. Proc. {\bf 30}, 011010 (2020).
\bibitem{Ozawa2016} R.~Ozawa, S.~Hayami, K.~Barros, G.-W.~Chern, Y.~Motome, and C.~D.~Batista, J. Phys. Soc. Jpn. {\bf 85}, 103703 (2016).
\bibitem{Shimizu2022PRB} K.~Shimizu, S.~Okumura, Y.~Kato, and Y.~Motome, arXiv:2201.03290, to be published in Phys. Rev. B.
\bibitem{footnote2} We calculate the ellipticity by $\xi_\eta=|\bold{u}^1_\eta|/|\bold{u}^2_\eta|$, where $\bold{u}^1_\eta$ and $\bold{u}^2_\eta$ are vectors defining the minor and major axes of the helical plane, respectively:
$\bold{u}^{1,2}_\eta=\bold{u}_\eta(\theta^*)$ with $\left.\frac{\partial}{\partial\theta}|\bold{u}_\eta(\theta)|\right|_{\theta=\theta^*}=0$, where $\bold{u}_\eta(\theta)=\mathrm{Re}\;\bold{S}_{\bold{Q}_\eta}\cos\theta+\mathrm{Im}\;\bold{S}_{\bold{Q}_\eta}\sin\theta$, to satisfy $|\bold{u}^1_\eta|\leq|\bold{u}^2_\eta|$ and $\bold{u}^0_\eta\cdot\left(\bold{u}^1_\eta\times\bold{u}^2_\eta\right)\geq0$. 
We obtain $\bold{u}^0_\eta$ by $\bold{u}^0_\eta\propto\mathrm{Re}\;\bold{S}_{\bold{Q}_\eta}\times\mathrm{Im}\;\bold{S}_{\bold{Q}_\eta}$. 
\bibitem{MacChesney1965} J.~B.~MacChesney, R.~C.~Sherwood, and J.~F.~Potter, J. Chem. Phys. {\bf 43}, 1907 (1965).
\bibitem{Takeda1972} T.~Takeda, Y.~Yamaguchi, and H.~Watanabe, J. Phys. Soc. Jpn. {\bf 33}, 967 (1972).
\bibitem{Oda1977} H.~Oda, Y.~Yamaguchi, H.~Takei, and H.~Watanabe, J. Phys. Soc. Jpn. {\bf 42}, 101 (1977).
\bibitem{Reehuis2012} M.~Reehuis, C.~Ulrich, A.~Maljuk, Ch.~Niedermayer, B.~Ouladdiaf, A.~Hoser, T.~Hofmann, and B.~Keimer, Phys. Rev. B. {\bf 85}, 184109 (2012).
\bibitem{Reimers1991} J.~N.~Reimers, A.~J.~Berlinsky, and A.-C.~Shi, Phys. Rev. B {\bf 43}, 865 (1991).
\bibitem{Okubo2011} T.~Okubo, T.~H.~Nguyen, and H.~Kawamura, Phys. Rev. B {\bf 84}, 144432 (2011).
\bibitem{Yambe2020} R.~Yambe and S.~Hayami, J. Phys. Soc. Jpn. {\bf 89}, 013702 (2020).
\bibitem{Mostovoy2005} M.~Mostovoy, Phys. Rev. Lett. {\bf 94}, 137205 (2005).
\bibitem{Azhar2017} M.~Azhar and M.~Mostovoy, Phys. Rev. Lett. {\bf 118}, 027203 (2017).
\bibitem{Hayashi2001} N.~Hayashi, T.~Terashima, and M.~Takano, J. Mater. Chem. {\bf 11}, 2235 (2001).
\end{thebibliography}
\end{document}